# Microsphere Super-resolution Imaging

## Zengbo Wang

*School of Electronic Engineering, Bangor University, Bangor, UK. LL57 1UT E-mail:*
*z.wang@bangor.ac.uk*

Recently, it was discovered that microsphere can generate super-resolution focusing beyond diffraction limit. This has led to the development of an exciting super-resolution imaging technique -*microsphere nanoscopy*- that features a record resolution of 50 nm under white lights. Different samples have been directly imaged in high resolution and real time without labelling, including both non-biological (nano devices, structures and materials) and biological (subcellular details, viruses) samples. This chapter reviews the technique, which covers its background, fundamentals, experiments, mechanisms as well as the future outlook.

## 1 Introduction

Optical microscopy is one of the most important scientific achievements in the history of mankind. The invention of the compound optical microscope by Hans & Zaccharis Janssen in 1590 and improvements by Galileo Galilei, Robert hook, Anthony Leeuwenhoek has revolutionized all aspects of science and technology, especially the life sciences when it became possible for researcher to see, for instance, the bacteria and blood cells.

In 1873, Ernst Abbe established the resolution limit of optical microscopes[1]: The minimum distance, d', between two structural elements to be imaged as two objects instead of one is given by d' = $\lambda$ / (2NA), where $\lambda$ is the wavelength of light and NA the numerical aperture of the objective lens. The physical root for resolution limit is related to optical diffraction and loss of evanescent waves in far-field; the evanescent waves carry high-frequency subwavelength spatial information of an object and decay exponentially with distance from the object. With white lights, optical microscope resolution is limited at about 200-250 nm. For about one hundred years, the Abbe criterion was considered the fundamental limit of optical microscope resolution.

One key step forward in beating diffraction limit was the invention of near-field scanning optical microscopy (NSOM) technique by D.W. Pohl and colleagues in 1984, allowing sub-wavelength optical imaging for the first time[2]. Here an image of a structure is constructed by scanning a physical tip with subwavelength aperture in the proximity (~tens of nanometres) of an illuminated specimen. Since the late 1990s, stimulated by the surge of nanophotonics, plasmonics and metamaterials, a number of new super-resolution microscopy/nanoscopy techniques have appeared which include metamaterial superlens[3] and STED (stimulated emission depletion microscopy)[4].

Metamaterial superlens, or Pendry superlens, was first theoretically proposed by British scientist John Pendry in 2000 which use a slab of NIM (negative-index medium) to enhance the evanescent waves, offers the possibility to restore the nanoscale information in the far-field and therefore a nearly perfect image can be reconstructed[3]. The 'perfect lens' concept was widely taken by the community and sparked a research surge in metamaterials and plasmonics. Several versions of plasmonic metamaterial superlenses, which follow Pendry's basic idea, have been developed and demonstrated by a number of groups and researchers across the world in the past decade[5-8]. It is important to note that in its currently most advanced version, called hyper-lens[6, 7], the evanescent waves are converted into propagating waves forming a magnified image of the sample on a distant screen, which is why one may think that it is far-field imaging. However, the projection to a distant screen does not change the fact that the hyperlens relies on the sample's near-field. Hence, in its current state of development, a hyperlens is not a far-field imaging device[6], but a non-scanning concept of employing the near-field. Because of the fundamental loss limit of plasmonic materials used in these lenses, and nanofabrication challenges, resolutions are still limited at 70~100 nm at a single visible wavelength after a decade's efforts by researchers across the world, and haven't been well appreciated for bio-imaging application.

Into the far-field domain, the super-resolution fluorescence microscopy techniques are the most successful developments in recent year. They are attained with two main approaches: spatially patterned excitation (STED, RESOLFTs, SSIM)[9] and single-molecule localization (STORM, PALM, FPALM)[9] of fluorescence molecules. The techniques are already applied on a large scale in major fields of the biological sciences, like cell biology, microbiology and neurobiology, and may revolutionize the entire biology and medicine fields in the future. The developers of these techniques (Erik Betzig, Stefan W. Hell and W. E. Moerner) were awarded the Nobel Prize in Chemistry in 2014. But since the techniques are fluorescence based, they cannot be used for imaging non-fluorescence samples, e.g. electronic devices and many viruses and sub-cellular structures that cannot be labelled using existing fluorophores. Meanwhile, we should notice the fact that these techniques are not based on high-resolution lens, but on fluorescent materials. The use of fluorescence may also change original function of studied bio-objects, and affect its dynamic processes as well. Therefore, there is still a great need to develop a super-resolution lens which can offer label-free, high-resolution imaging of any samples.

Another far-field super-resolution technique demonstrated recently is the 'optical superoscillatory lens' by Zheludev's group in Southampton[10]. The basic idea, which has root connection with Toraldo di Francia's proposal in 1956[11], is to use a carefully-designed phase-mask to modulate the beam to achieve a super-resolution spot in the far-field, through constructively and destructively interference without evanescence waves being involved. The key disadvantage in the technique is the appearance of giant sidelobes near to the central spot, which seriously affected the practical adoption of the technology.

In 2011, a new technique -*microsphere nanoscopy*- was developed by the present author and his colleagues[12]. The technique uses micro-sized spheres as super-resolution lens (*'microsphere superlens'*) to magnify underlying objects before projecting them into a conventional microscope's objective lens. The technique is label-free in nature, offering a remarkable resolution of 50 nm under white lights. These features are unique and attractive for achieving low-intensity high-resolution imaging of almost any nanoscale objects[12-17].

In addition, a number of other super-resolution techniques and proposals exist in the literature, including structured illumination, Maxwell fisheyes, scattering lens and time-reversal imaging. More details on them can be found in an interesting magazine article written by J. Cartwright[18].

## 2 Microsphere super-resolution focusing

Unlike metamaterial superlens and hyperlens, which use metal to amplify evanescent waves to achieve superresolution[5], the microsphere superlens uses a different strategy in achieving the focusing super-resolution - *'photonic nanojet (PNJ)'*. The PNJ is attained via light scattering by micro-sized particles (spheres, cylinders, typical diameter between 1- 50 μm) [19, 20]. The early paper relating to this effect can be dated back to year 2000: Lu and Luk'aynchuk et al. first demonstrated that enhanced optical near-fields of a 500-nm silica sphere can be used as superlens for subwavelength structuring of silicon surface[21]. Since, there has been a strong continuing interest on the technique and many progresses have been made. Theoretical studies related to the topic also grew rapidly[22]. Without knowing the initial work by Lu and Luk'yanchuk, in 2004 Chen et al. coined a new term *'photonic nanojet'* for dielectric particle super-resolution focusing at the shadow side of the particle[19], which is now widely adapted and used in optics community[19, 23-34].

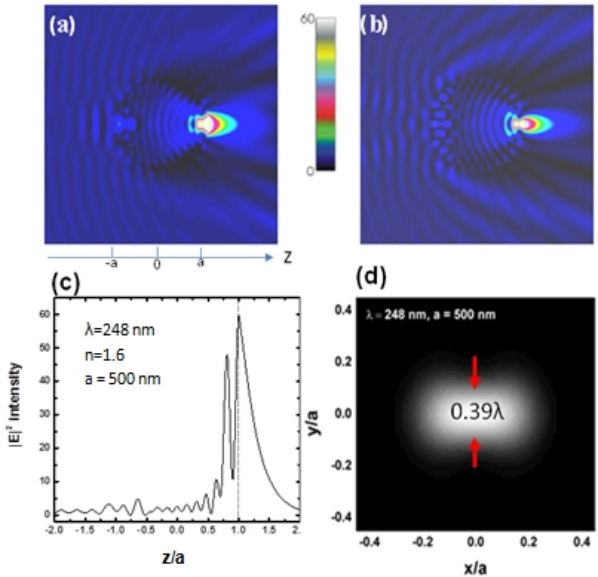

**Fig. 1** Photonic nanojet sspatial Intensity distribution, $I = |E|^2$, inside and outside the 1.0 μm PS particle, illuminated by a laser pulse at $I$ =248 nm, and (a) polarization parallel and (b) perpendicular to the image plane. The maximum intensity enhancement in calculations is about 60 for both regions. (c) Shows the intensity along z-axis. $z = 1.0$ is the position under the particle. (d) Super-resolution spot at z=a, the tangent plane right under the particle.

The key properties of PNJ include[20, 29]:

    (1) The transverse beam wide of the nanojet can reach λ/2n [35], where λ is light wavelength and n the refractive index of particle. In case of a polystyrene

particle with n=1.6, the resolution limit is about $\lambda/3.2 \approx 0.313\ \lambda$ following the literature. Here, one should note the super-resolution focus is located outside of particle and in the air zone which is accessible for nanoscale optical applications.

(2) It is a non-resonant phenomenon that can appear for a certain range of the diameter d of the dielectric microsphere or microcylinder from $2\lambda$ to more than $40\lambda$ if the refractive index contrast relative to the background medium is less than about 2:1.

In early stage of the PNJ development, the main application is in surface nanopatterning and nanofabrication[21, 36-43]. Typically a pulsed nanosecond or femtosecond laser is applied as laser source, the laser radiation goes through particles and a near-field PNJ focus is formed next to the particle exit surface. An example of such optical near-field distributions calculated by Mie theory is shown in Fig. 1 for a 1.0 μm Polystyrene (PS) particle illuminated by a laser pulse at $\lambda = 248$ nm.

Figure 1 reveals basic characteristics of PNJ of a microsphere lens. The electric field is greatly enhanced in the near-field zone under the particle, but quickly decays along light propagation z-direction (almost exponentially as a character of near-field optics), from 59.6 at $z = a$ to 1.57 at $z = 2a$. Within the $z = a$ tangential plane, the distribution of laser intensity shows an elongated profile, whose long axis aligns with incident beam polarization while the super-resolution is observed in the cross direction. A resolution of 0.39 $\lambda$ has been achieved in presented case, which surpasses the 0.61$\lambda$ (Rayleigh criteria) or 0.5$\lambda$ (Sparrow criteria) diffraction limit. The elliptical profile of near-field intensity has a physical root associated with the radial component of the electric field, $E_r$, which decays with $r$ as $E_r \propto 1/r^2$, in near-field zone. It quickly decays to zero in far field zone($r \geq 1$) [44]. In other words, scattering wave in the far field is transverse but contains both transverse and longitudinal components in near-fields.

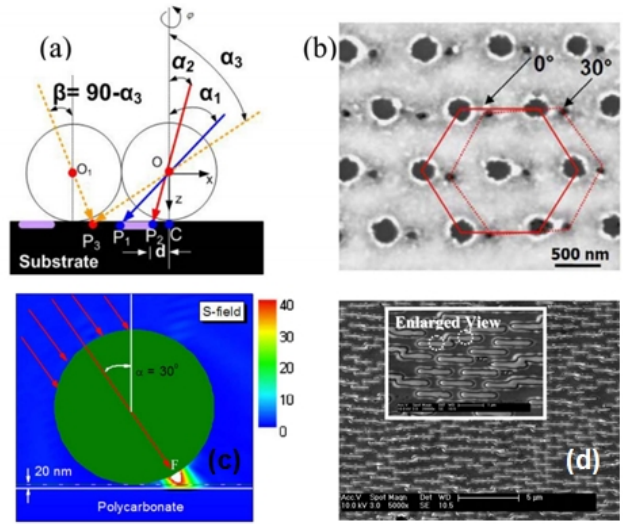

**Fig. 2** Schematic diagram of (a) experimental configuration for direct laser writing of nanopatterns on substrate surface, 1um SiO2 particles were used. (b) SEM image of two hexagonal array ablated by single KrF 248 nm wavelength laser at angle 0 (vertical incident), fluence 6.5 mJ/cm2 and 30-deg, fluence 1.0 mJ/cm2, respectively. Small hole is 80 nm. (c) Calculated field intensity at 45-deg incidence angle. (d) Nano-lines, length 1400 nm, width 360 nm fabricated on substrate surface.

An example of laser surface nanopatterning using microsphere superlens was given in Fig. 2. Here, a laser beam was angularly scanned (Fig. 2a), causing the focus spot of each microsphere lens incident on different locations of the sample (Fig. 2c). A resolution of 80 nm has been achieved in the experiments – the small nanoholes in Fig.2b. By continuously tuning the incident angles, different nanopatterns (Fig. 2b, 2d) can be fabricated on sample surface over a large surface area, directly without a photo-mask. The technique is promising for next-generation low-cost, large-area laser nanomanufacturing.

## 3 Microsphere nanoscopy super-resolution imaging

### 3.1 Widefield microsphere nanoscopy

Imaging might be considered as an inverse process of nanofabrication discussed above. In 2011, we successfully demonstrated that, by simply putting microsphere superlens on top of nanoscale samples, and using a conventional microscope to look through these mini lenses, we were able to see nano-features down to 50 nm with a white lighting source, directly without any complex labelling process[12, 16].

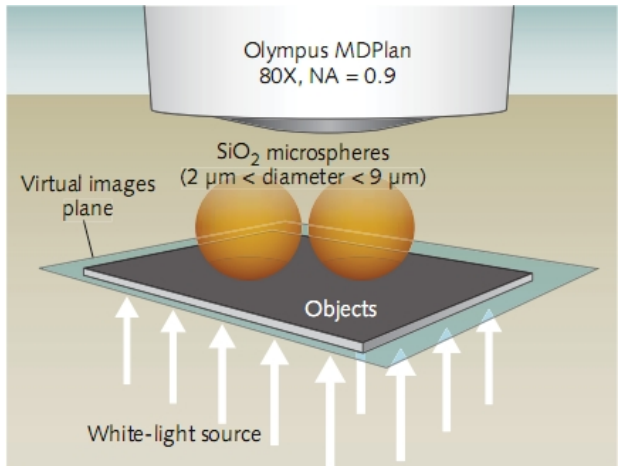

**Fig. 3** A schematic shows a white-light microsphere nanoscope (a microsphere superlens integrated with a classical widefield optical microscope) with $\lambda/8$ imaging resolution. The spheres collect the near-field object information and form virtual images that are then captured by the conventional lens.

The technique is based on the use of superlensing microspheres made from ordinary materials such as silicon dioxide ($SiO_2$) and polystyrene. Figure 3 shows the typical experimental setup which we called it *'microsphere nanoscope'*. The microspheres are placed on top of the object's surface by self-assembly. The as-received $SiO_2$ microsphere suspension (from Bangs Laboratories in Fishers, IN) is diluted and applied to the imaging samples by drop or dip coating and the samples are left to dry in air. A halogen lamp with a peak wavelength of 600 nm is used as the white-light illumination source. These microspheres function as superlenses - *microsphere super lenses*-that collect the underlying near-field object information and magnify it (forming virtual images that keep the same orientation as the objects in the far-field) before it is projected to an 80X Olympus (Essex, England) objective lens ($NA = 0.9$, model MDPlan) of an Olympus microscope (model MX-850). The combination of microsphere superlenses and the objective lens forms a compound-imaging lens system. In the reflection mode, the white-light source will be incident from the top, opposite to the light source at the bottom in the transmission mode.

We have captured clear images of sub-diffraction-limited features of nanoscale objects using a microsphere nanoscope in either transmission or reflection mode. For example, 30-nmthick chrome-film diffraction gratings with 360-nm-wide lines spaced 130 nm apart on fused silica substrates were imaged in transmission mode (see Fig. 4b top half). The virtual image plane was 2.5 µm beneath the substrate surface, and only those lines with microsphere particles on top of them were resolved. The lines without the particles mix together and form a bright spot that cannot be directly resolved by the optical microscope because of the diffraction limit. For the visible wavelength 400 nm, the best diffraction-limited resolution is estimated to be 215 nm in air using the vector theory of Richards and Wolf [45], and 152 nm when taking the solid-immersion effect of a particle into account. For the main peak of a white-light source at 600 nm, the limits are 333 nm in air and 228 nm with solid-immersion effect, respectively. Here, one should also note that the focal planes for the lines with and without particles on top are different.

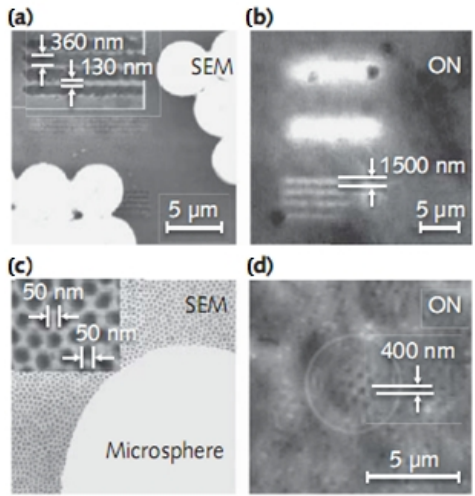

**Fig. 4** Two examples show a microsphere superlens imaging in transmission mode. For a diffraction grating with 360-nm-wide lines spaced 130 nm apart (a; top-left image taken by scanning electron microscope), the optical nanoscope (ON) image (b) shows that the lines are clearly resolved. The magnified image corresponds to a 4.17X magnification factor. For a gold-coated fishnet membrane sample imaged with a microsphere (size 4.7 μm; borders of two spheres are shown by white lines) superlens. The nanoscope clearly resolves the pores that are 50 nm in diameter and spaced 50 nm apart (c; SEM image). The size of the optical image between the pores within the image plane is 400 nm (d; ON image) and corresponds to a magnification factor of approximately 8X.

In another example (Fig. 4 bottom-half images), a fishnet 20-nmthick gold-coated anodic aluminium oxide (AAO) membrane fabricated by two-step anodizing in oxalic acid (0.3 mol/l) under a constant voltage of 40 V is imaged with 4.7-μm-diameter microspheres. The membrane pores are 50 nm in diameter and spaced 50 nm apart. The microsphere nanoscope resolves these tiny pores well beyond the diffraction limit with a resolution between λ/8 (λ = 400 nm) and λ/14 (λ = 750 nm) in the visible spectral range. It is important to note that the magnification in this case is around 8X—almost two times that of the earlier grating example, implying that the performance of the microsphere superlens is affected by the near-field interaction of the sphere and the substrate.

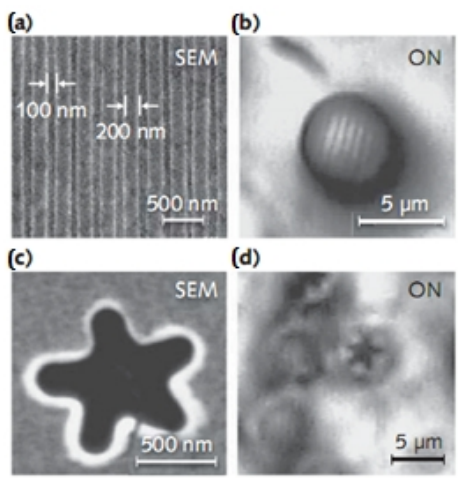

**Fig. 5** Microsphere nanoscope in reflection mode images a commercial Blu-ray DVD disk. The 100-μm-thick transparent protection layer of the disk was peeled off before using the microsphere (size 4.7 μm). The sub-diffraction-limited 100 nm lines (a; SEM image) are resolved by the microsphere superlens (b; ON image). In a second example, reflection-mode imaging of a star structure made on a DVD disk thin film (c; SEM image) is clearly discerned (d; ON image)— including the complex shape of the star and its 90-nm-diameter corners.

Since self-assembled particles are easy to spread over a large surface area with each particle working as a superlens, the images produced by each particle can be stitched together to form a large image; that is, a hexagonal array of particles functions as a superlens array covering a large area.

Microsphere superlens imaging can also be accomplished in the reflection mode using halogen light illumination; in fact, the sub-diffraction-limited lines in a Blu-ray DVD disk (200-nm-wide lines with 100 nm separation) are clearly imaged with 4.7-μm-diameter microspheres (see Fig. 5). The microsphere superlens can also discern the shape of a star structure made on an antimony-tellurium (SbTe) DVD disk. The complex shape of the star, including its 90-nm-diameter corners, was clearly resolved by the microsphere superlens in reflection mode.

### 3.2 Confocal microsphere nanoscopy

In widefield microscope, the imaging contrasts are often low and unsatisfactory due to the presence of out-of-focus light in the final image. To enhance the contrast, considerable efforts were often put to optimize the microscope lighting condition and imaging software settings during imaging. In contrast to widefield microscopy, confocal microscopy techniques generally have much better optical contrast and improved resolution; this is achieved by placing a tiny pinhole before the detector to eliminate the out-of-focus light in the final image. We also evaluated the imaging performance of microsphere-assisted confocal nanoscopy by integrating microsphere superlens with a 405 nm laser scanning confocal microscope (Olympus OLS4100), the results show that resolution can increase to 40 nm (Yan et al. claimed 25 nm resolution when using same setup[46], but we were not able to resolve 25 nm nano-line sample in our tests). This is because confocal microscope adopts a tiny pinhole before the detector which rejects out-of-focus light in final image, which theoretically could boost resolution by a factor of ~1.5 (theoretically 50/1.5~33 nm).

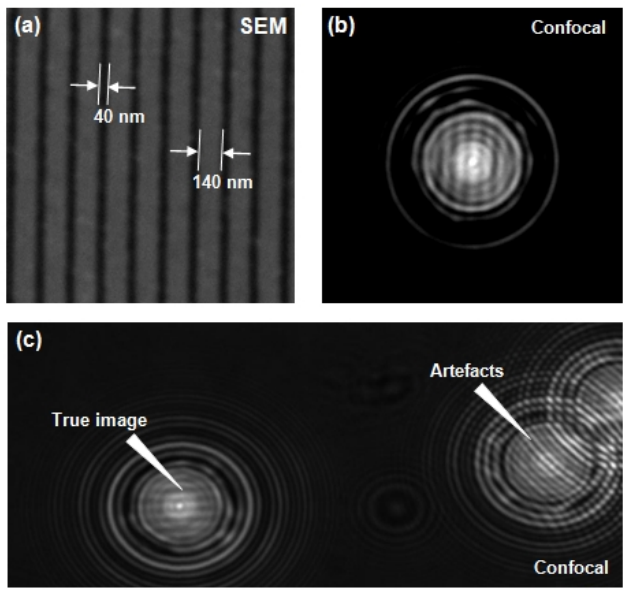

**Fig.6** (a) SEM image of the test sample: 40-nm nano lines with seperation distance of 140 nm, structructed on a 40-nm-thick gold film coated on a glass surstrate. (b) Corresponding confocal microsphere nanoscope image of the sample, taken with 50x/NA0.95 objective lens, LEXT4100 laser confocal micrsocope with wavelength 405 nm (c) Artefacts due to two-particle interference versus true pattern under single particle.

Figure 6 illustrates SEM and corresponding confocal imaging results, in which the 405 nm laser scanning confocal microscope was coupled with 4.7-μm-sized SiO2 microsphere deposited on sample surface. Here the sample contains 40 nm line features (dark strips) spaced 140 nm apart (bright gold strips). We tested imaging in both widefield and confocal modes. The structures cannot be discerned by the widefield nanoscope (picture not shown); whereas in confocal scheme, both 40 nm and 140 nm features can be clearly resolved. Notice here imaging was obtained with isolated single microspheres only. The particle viewing window and magnification factor is about 3.3 μm and M~3 (both averaged values), respectively.

Another finding in confocal imaging is the appearance of multiple concentric rings in the confocal imaging (Fig. 6c). These rings are resulting from near-field interactions of particle and substrate under a coherent laser illumination. In contrast, because of the use of incoherent light source, such issue is less obvious in the widefield. These rings degrade imaging quality, which may pose a practical limit on the minimum feature we can see in confocal imaging. In principle, the rings could be pre-calculated and subtracted from the final image through image processing.

Attentions should be paid to the artefacts that might be wrongly interpreted as object image. They result from coherent light interference between multiple particles, especially those closely positioned. One such example was illustrated in Fig. 6(c): on its right, one can see a double-particle system producing strong dots and grating artefacts; these images are artificial and not representing real features under the particles. One may check this by rotating the sample during experiments. As sample rotates, these artefacts didn't follow the sample rotation. Fortunately, for isolated particles, we can still see true image of the objects through them, as evidenced in Fig. 6(c) left part. The artefacts issue is less obvious in widefield nanoscopy system where an incoherent lighting source is often used [47].

### 3.3 Solid-Immersion microsphere superlens

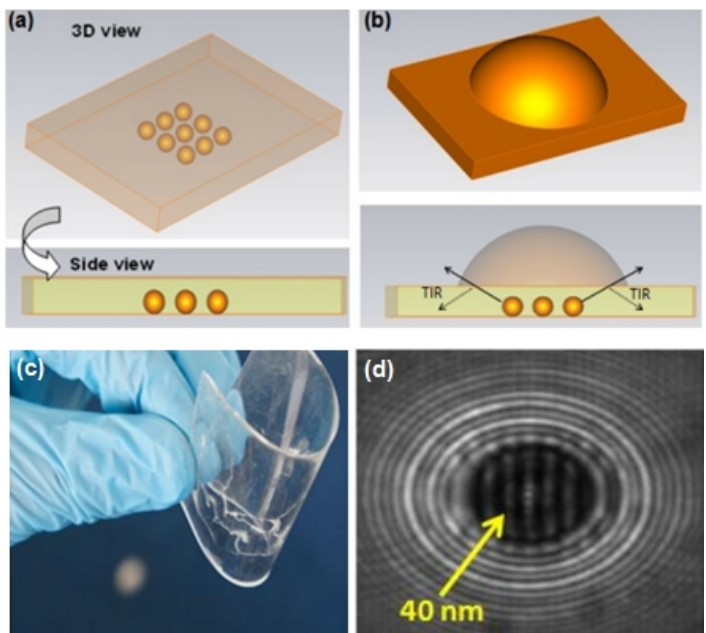

Fig.7 Solid-immersion microsphere superlens, (a) planar design with high-index microsphere embedded in transparent layer, (b) hemisphere-top design which reduces total internal refection (TIR) effect of planar surface, (c) lab prototype produced at Bangor University, and (d) imaging example showing 40 nm lines can be clearly imaged.

Practical applications in microscopy require control over the positioning of the microspheres for scanning operation. We proposed two approaches to solve this problem. In one proposal, the movement of microsphere was carried out with a fine glass micropipette attached to the microsphere[14]. The second proposal, which was recently patented[48], was to use high-index microspheres (TiO$_2$ or BaTiO$_3$) embedded into a transparent host material (such as PMMA and PDMS). This concept was also explored by Darafsheh and Astratov et al. in the US[49, 50].

Figure 7 shows two designs of coverslip-form solid-immersion microsphere superlens we have proposed, a flat-top design (Fig.7a), and a hemisphere-top design (Fig.7b), both having a thickness similar to standard coverslip which is small enough for the superlens to be directly inserted into the gap between conventional microscope's objective lens and sample. The hemisphere-top curved surface in the design Fig.7(b) is used to reduce the effect of total internal reflection from a flat surface.

The advantages of solid-immersion design include:

(1) Sphere lenses are now reusable and the whole lens can be easily positioned and manipulated according to the user needs.

(2) By attaching to a nano-stage, scanning operations are now possible.

(3) Higher resolution can be achieved; this is because in this design the solid immersion mechanism contributes to the resolution enhancement, as the working wavelength in the near-field region is scaled down by a factor of n (refractive index of solid immersion material). As shown in Fig. 7(d), lines with size 40 nm can be clearly imaged with the solid-immersion superlens.

(4) The new lens is also less sensitive to the environment change owing to encapsulation which separates microspheres and the environment. The solid-immersion design has great commercial potential which will be discussed in the outlook session below.

## 3.4 Other developments

There are number of interesting developments by other groups. Hao and Liu et al. demonstrated that imaging contrast in the widefield microsphere nanoscope could be improved by partly immersion of microspheres in liquids[51]. They also proposed and demonstrated a nice variant of the technique, employing a *microfiber* to evanescently illustrate the specimen while simultaneously imaging them at a 75-nm resolution[52]. Imaging contrast in this case was greatly improved owing to the limited illumination depth (typical <200 nm) of the evanescent waves; sharp and clear images of nanostructures have been achieved[52]. This idea may worth further exploration for sub-50 nm resolution imaging in the future, by extending the evanescence wave illumination approach to microspheres; either prism-style or objective-style TIRM (Total Internal Reflection microscopy) setup can be used [53]. Another important development, of particular importance for biological applications is the demonstration of super-resolution imaging with microspheres inside aqueous environments; here higher-index microspheres[54, 55] were chosen to match the background liquid medium. Darafsheh et al. also demonstrated a series of advantages of microsphere-assisted imaging over confocal and solid immersion lens microscopies, including intrinsic flexibility, better resolution, higher magnification, and longer working distances. They discerned minimal feature sizes of ~50-60 nm in nanoplasmonic arrays. Their results are in consistent with our findings of 50 nm resolution in most cases. On the other hand, Vlad et al. studied the imaging performance of thermally reshaped polymer microspheres and addressed the issue on virtual image plane selection[47].

In biological fields, Yang et al. imaged biological samples and demonstrated that the shape of subcellular structures like centrioles, mitochondria and chromosomes can be clearly resolved through the microsphere nanoscope. The imaging method is also used to identify the expression of the specific mitochondrial membrane protein MTCO1[17]. Li et al. applied the technique to image adenovirus without labelling[15].

# 4 Super-resolution mechanism

## 4.1 Decoupling of evanescent waves

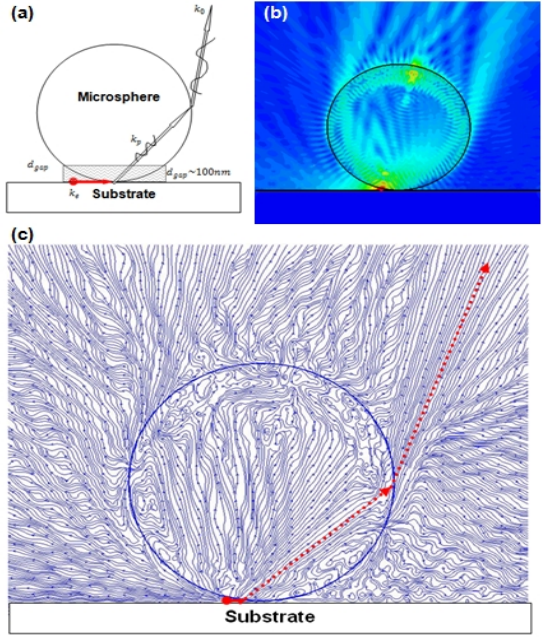

Fig.8 Decoupling of high spatial frequency evanescent waves $k_e$ by the particle-substrate system in microsphere nanoscope. Incident wavelength 600nm, particle size 4.7 µm, refractive index 1.46.

Fine details of nanostructures are carried by the high spatial frequency evanescent waves localized in the optical near fields, which exponentially decay from the source. In NSOM, a tiny local probe (aperture) was brought into the optical near fields that interacts with the evanesced waves. The probe functions as an optical antenna that converts localized energy into radiating waves which propagates into the far-field. This basic concept applies to our technique as well. In microsphere nanoscopy, microspheres are in direct contact with objects, which forms a Particle-on-Surface (POS) system, as illustrated in Fig.8(a). It is this POS system that effectively interacts with the near-field evanescent waves, decouples and turns them into propagating wave that would reach the objective lens in the far-field. Like in NSOM, such interaction are quite complicated and only takes place in the proximity of the interface of particle and substrate, which was indicated by the shadow zone in Fig. 8(a). The characteristic thickness of the evanescent waves scattering zone, $d_e$ , could be decided by:

$$d_e = r - \sqrt{r^2 - \frac{x^2}{4}}$$

where $r$ is the particle radius and $x$ the size of sample viewing window, which was given by:

$$x = \frac{xp}{M}$$

with $xp$ represents size of particle viewing window and $M$ the magnification factor.

Both quantities are directly measureable through experiments. Taking imaging results in Fig.6 as example, we have

$$x \cong 3.3/3 = 1.1 \, \mu m \; ; \; d_e \cong 65 \, nm$$

Here, the viewing window size, $x$, is about *one quarter* of the particle diameter, and the evanescent waves decaying length is on the order of *~100 nm,* which only accounts for ~2% of the particle diameter. Notably, this tiny interface layer is of critical importance for the microsphere nanoscopy technique; any method that could effectively affect the evanescent waves in this region would play an important role for further development of the technique. Here we discuss two possible proposals. The first relates to the creation of a thin higher-index immersion layer at the sample surface which fills the curved meniscus of POS gaps ($n_{gap} > 1$); for example by adding a thin high-index film with thickness of a few hundred nanometres. In this case, since

$$k_z = i \sqrt{k_x^2 + k_y^2 - n_{gap}^2 . k_0^2}$$

, it will deceases when higher-index material presented in the gap zone; the evanescent wave zone will be extended that would strengthen the coupling between evanescent waves and microspheres [51]. Another approach is to learn from Pendry's superlens idea: using a slab of negative-index medium to amplify the surface evanescent wave signal so that coupling to microspheres could be enhanced [3]. Practically this could be done by coating plasmonic films on particle and sample surface; notice in this case both material and thickness of the film have to be properly chosen to match the illuminating wavelength and sample properties [8]. This proposal has interesting connection with recently purposed 'adiabatic lens' by Cang et al. [56]

To illustrate the evanescent waves scattering process by POS system, numerical simulation was carried for a point dipole source located within the gap of POS system. Such calculation is highly intensive, it took about 10 days on a 75-GB memory, 8 Cores 3.2 GHZ CPU PC, which suggests an analytical solution is highly desirable in the future development. From the results shown in Fig. 7, the speciality of this POS system can be observed: the microspheres can effectively pickup large-angle surface waves, bending them and projecting them into the far-field.

### 4.2 Super-resolution illumination and resolution limit

It is noted that theoretically calculated smallest 'photonic nanojet' spot size, roughly between 100 nm and 150 nm when neglecting substrate effect, cannot explain the observed experimental resolution of 50 nm[57-59]. In practise, the sample is positioned away from the focusing plane, but at a plane contacting the microsphere bottom surface, rendering an out-of-focus virtual imaging mode. This means the sample surface is not illuminated by single focused 'photonic nanojet' spot, but a two dimensional landscape fields that can contain higher spatial resolution spots. Figure 9 demonstrates such philosophy. Despite the focus spot is larger than 100 nm, the illuminating spot at sample plane show sub-100 nm scale focusing which will contribute to the high resolution imaging. The estimated theoretical limit of microsphere nanoscopy technique is around 20 to 30 nm[12]. Calculations neglecting substrate and illumination conditions may reach inaccurate conclusions, and cares should be taken in the modelling processes[57].

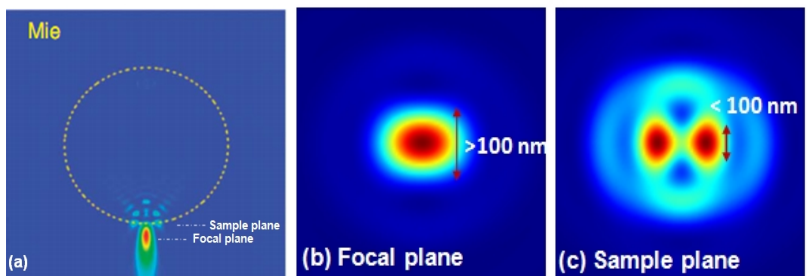

Fig.9 Illustration of super-resolution illumination concept. (a) Filed distribution in incident plane (b) focus spot at focal plane, typically focus spot > 100 nm according, and (c) reduced illumination spot in sample plane.

## 5 Outlooks

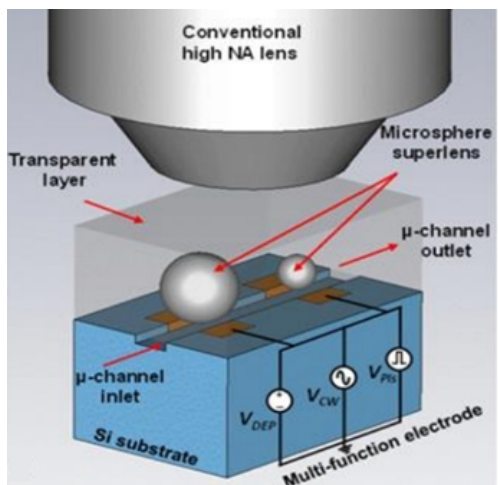

Fig.10 Proposed new on-chip microfludic nanoscope formed by integrating nanoscope with microfludics, providing a unique platform for real-time high resolution direct imaging and analysis of bio-samples, including viruses.

Compared to other super-resolution techniques, the micropshere nanoscopy technique has a number of distinct features– simple, easy to implement, label-free, high-resolution and support white-lighting imaging. Also, because microsphere superlenses are small in size, they are naturally excellent for being integrated with other microsystems, e.g. microfludics, to form new function on-chip devices. An example proposal is shown in Fig.10, where microsphere superlenses are encapsulated, aligned and positioned right on top of a microfludic channel in which biological objects such as cells/viruses can flow through the channel, its high-resolution image will be projected in real-time to the outside world. Using an electro-isometric flow it is possible to drive the bio-species along the channel, and use dielectrophoresis force to stop and trap the cells/viruses at particular locations. This will allow researchers, for example, to study the real-time response of cells/viruses when they interact with medicines.

Commercialization of technology is another direction to go. Recent development of commercially viable superlens prototype product at Bangor University is an important achievement (Fig.7). More efforts are currently paid to optimize the product. The product can effectively transform a regular microscope into a nanoscope (offering 100 nm resolution in wide field and 50 nm resolution in confocal mode). The superlens resembles a standard coverslip glass design so it is naturally compatible with all existing microscopes used in teaching and research. It is envisaged that someday every microscope user will have the superlens product in their hand for daily use of microscopes. Besides imaging, the superlens product can find applications in nanofoucsing, nanolithography, nano-solar energy concentrator, nanochemistry, nanomedicine areas as well.

## References

‡ white lights are low-intensity beams, free of phototoxic and photobleaching problems as often encountered in laser beams; they are the most commonly used lighting source in light microscopes.
1.      E. Abbe, *Archiv Microskop. Anat.*, 1873, **9**, 413.
2.      D. W. Pohl, W. Denk and M. Lanz, *Applied Physics Letters*, 1984, **44**, 651-653.
3.      J. B. Pendry, *Phys. Rev. Lett.*, 2000, **85**, 3966.
4.      S. W. Hell and J. Wichmann, *Opt Lett*, 1994, **19**, 780-782.
5.      X. Zhang and Z. W. Liu, *Nature Materials*, 2008, **7**, 435-441.
6.      Z. W. Liu, H. Lee, Y. Xiong, C. Sun and X. Zhang, *Science*, 2007, **315**, 1686-1686.
7.      Z. Jacob, L. V. Alekseyev and E. Narimanov, *Optics Express*, 2006, **14**, 8247-8256.
8.      N. Fang, H. Lee, C. Sun and X. Zhang, *Science*, 2005, **308**, 534-537.
9.      B. Huang, M. Bates and X. Zhuang, *Annual review of biochemistry*, 2009, **78**, 993-1016.
10.     E. T. Rogers, J. Lindberg, T. Roy, S. Savo, J. E. Chad, M. R. Dennis and N. I. Zheludev, *Nat Mater*, 2012, **11**, 432-435.
11.     G. T. d. Francia, *Nuovo Cimento, Suppl.*, 1952, **9**, 426-435.
12.     Z. Wang, W. Guo, L. Li, B. Luk'yanchuk, A. Khan, Z. Liu, Z. Chen and M. Hong, *Nat Commun*, 2011, **2**, 218.
13.     A. Darafsheh, N. I. Limberopoulos, J. S. Derov, D. E. Walker and V. N. Astratov, *Applied Physics Letters*, 2014, **104**, -.
14.     L. A. Krivitsky, J. J. Wang, Z. B. Wang and B. Luk'yanchuk, *Sci Rep-Uk*, 2013, **3**.
15.     L. Li, W. Guo, Y. Yan, S. Lee and T. Wang, *Light: Science & Applications*, 2013, **2**, e104.
16.     Z. B. Wang and L. Li, *Laser Focus World*, 2011, **47**, 61-64.
17.     H. Yang, N. Moullan, J. Auwerx and M. A. M. Gijs, *Small*, 2013, DOI: 10.1002/smll.201302942, n/a-n/a.
18.     J. Cartwright, *Physics world*, 2012, **25** 29-34.
19.     Z. G. Chen, A. Taflove and V. Backman, *Optics Express*, 2004, **12**, 1214-1220.
20.     Z. B. Wang, Normal, National University of Singapore, 2005.
21.     Y. F. Lu, L. Zhang, W. D. Song, Y. W. Zheng and B. S. Luk'yanchuk, *JETP Lett.*, 2000, **72**, 457-459.
22.     Z. B. Wang, J. Naveen, L. Li and B. S. Luk'yanchuk, *P.I.Mech. Eng. C-J. Mec.*, 2010, **224**, 1113-1127.
23.     V. Yannopapas, *Optics Communications*, 2012, **285**, 2952-2955.
24.     D. McCloskey, J. J. Wang and J. F. Donegan, *Optics Express*, 2012, **20**, 128-140.
25.     M. S. Kim, T. Scharf, S. Muhlig, C. Rockstuhl and H. P. Herzig, *Applied Physics Letters*, 2011, **98**.

26.   D. Grojo, L. Charmasson, A. Pereira, M. Sentis and P. Delaporte, *J Nanosci Nanotechno*, 2011, **11**, 9129-9135.
27.   C. M. Ruiz and J. J. Simpson, *Optics Express*, 2010, **18**, 16805-16812.
28.   Y. E. Geints, E. K. Panina and A. A. Zemlyanov, *Optics Communications*, 2010, **283**, 4775-4781.
29.   A. Heifetz, S. C. Kong, A. V. Sahakian, A. Taflove and V. Backman, *Journal of Computational and Theoretical Nanoscience*, 2009, **6**, 1979-1992.
30.   P. Ferrand, J. Wenger, A. Devilez, M. Pianta, B. Stout, N. Bonod, E. Popov and H. Rigneault, *Optics Express*, 2008, **16**, 6930-6940.
31.   M. Gerlach, Y. P. Rakovich and J. F. Donegan, *Optics Express*, 2007, **15**, 17343-17350.
32.   X. Li, Z. G. Chen, A. Taflove and V. Backman, *Optics Express*, 2005, **13**, 526-533.
33.   A. V. Itagi and W. A. Challener, *Journal of the Optical Society of America a-Optics Image Science and Vision*, 2005, **22**, 2847-2858.
34.   Z. G. Chen, A. Taflove and V. Backman, *Ieee Antennas and Propagation Society Symposium, Vols 1-4 2004, Digest*, 2004, 1923-1926.
35.   H. Guo, Y. Han, X. Weng, Y. Zhao, G. Sui, Y. Wang and S. Zhuang, *Opt Express*, 2013, **21**, 2434-2443.
36.   W. Guo, Z. B. Wang, L. Li, Z. Liu, B. S. Luk'yanchuk, P. L. Crouse and D. J. Whitehead, 24-28 April 2007, University of Vienna, Vienna, Austria, 2007.
37.   W. Guo, Z. B. Wang, L. Li, D. J. Whitehead, B. S. Luk'yanchuk and Z. Liu, *Appl. Phys. Lett.*, 2007, **90**, 243101.
38.   G. X. Chen, M. H. Hong, Y. Lin, Z. B. Wang, D. K. T. Ng, Q. Xie, L. S. Tan and T. C. Chong, *Journal of Alloys and Compounds*, 2008, **449**, 265-268.
39.   W. Guo, Z. B. Wang, L. Li, Z. Liu, B. Luk'yanchuk and D. J. Whitehead, *Nanotechnology*, 2008, **19**, 455302.
40.   L. Li, W. Guo, Z. B. Wang, Z. Liu, D. J. Whitehead and B. S. Luk'yanchuk, presented in part at the The 1st International Conference on Nanomanufacturing (nanoMan2008), Singapore, July 14-16, 2008, 2008.
41.   Y. Zhou, M. H. Hong, J. Y. H. Fuh, L. Lu, B. S. Lukyanchuk and Z. B. Wang, *Journal of Alloys and Compounds*, 2008, **449**, 246-249.
42.   A. Khan, Z. Wang, M. A. Sheikh and L. Li, *International Journal of Manufacturing, Materials, and Mechanical Engineering (IJMMME)*, 2011, **1**, 9.
43.   A. Khan, Z. Wang, M. A. Sheikh, D. Whitehead and L. Li, *Appl. Surf. Sci.*, 2011, **258**, 774-779.
44.   M. Born and E. Wolf, *Principles of optics*, Cambridge University Press, UK, 7 edn., 1999.
45.   B. Richards and E. Wolf, *Proc. R. Soc. London Ser. A*, 1959, **253**, 358.
46.   Y. Yan, L. Li, C. Feng, W.Guo, S.Lee and M.H.Hong, *Acs Nano*, 2014, **8**, 1809-1816.
47.   A. Vlad, I. Huynen and S. Melinte, *Nanotechnology*, 2012, **23**.
48.   Z. Wang, *PCT/GB2014/052578 (priority date: 2013-AUG-23)*, 2014.
49.   K. W. Allen, N. Farahi, Y. Li, N. I. Limberopoulos, D. E. Walker, A. M. Urbas, V. Liberman and V. N. Astratov, *Annalen der Physik*, 2015, **527**, 513-522.
50.   V. N. Astratov and A. Darafsheh, *US patent application 2014/0355108 A1 published on December 4, 2014*, 2014.
51.   X. Hao, C. Kuang, X. Liu, H. Zhang and Y. Li, *Applied Physics Letters*, 2011, **99**, 203102.
52.   X. Hao, X. Liu, C. Kuang, Y. Li, Y. Ku, H.Zhang, H. Li and L. Tong, *Appl. Phys. Lett.*, 2013, **102**, 013104.
53.   P. R. Selvin and H. Taekjip, *Single-molecule techniques: a laboratory manual*, Cold Spring Harbor Laboratory Press, 2007.
54.   A. Darafsheh, G. F. Walsh, L. Dal Negro and V. N. Astratov, *Applied Physics Letters*, 2012, **101**.
55.   S. Lee, L. Li, Z. Wang, W. Guo, Y. Yan and T. Wang, *Appl Opt*, 2013, **52**, 7265-7270.
56.   H. Cang, A. Salandrino, Y. Wang and X. Zhang, *Nat Commun*, 2015, **6**, 7942.
57.   Y. Duan, G. Barbastathis and B. Zhang, *Opt Lett*, 2013, **38**, 2988-2990.
58.   S. Lee, L. Li, Y. Ben-Aryeh, Z. Wang and W. Guo, *Journal of Optics*, 2013, **15**, 125710.
59.   S. Lee, L. Li and Z. Wang, *Journal of Optics*, 2014, **16**, 015704.